\documentstyle[twocolumn,pre,psfig,floats,aps]{revtex}
\psfull 
\begin{document}

\twocolumn[\hsize\textwidth\columnwidth\hsize\csname@twocolumnfalse\endcsname
\title{Order parameter model for unstable multilane traffic flow}
\author{Ihor A. Lubashevsky\thanks{e-mail: ialub@fpl.gpi.ru}}
\address{Physics Department, Moscow State University, 119899 Moscow, Russia}
\author{Reinhard Mahnke\thanks{e-mail: reinhard.mahnke@physik.uni-rostock.de}}
\address{Fachbereich Physik, Universit\"at Rostock, D--18051 Rostock,
Germany}
\date{\today}
\maketitle

\begin{abstract}
We discuss a phenomenological approach to the description of unstable
vehicle motion on multilane highways that explains in a simple way the
observed sequence of the phase transitions ``free flow $\leftrightarrow$
synchronized mode $\leftrightarrow$ jam'' as well as the hysteresis in
these transitions.
We introduce a new variable called order parameter that accounts for
possible correlations in the vehicle motion at different lanes. So, it is
principally due to the ``many-body'' effects in the car interaction in
contrast to such  variables as the mean car density and velocity being
actually the
zeroth and first moments of the ``one-particle''
distribution function. Therefore, we regard the order parameter as an
additional independent state variable of traffic flow. We assume that
these correlations are due to a small group of ``fast'' drivers and taking
into account the general properties of the driver behavior

formulate a governing equation for the order parameter.
In this context we analyze the instability of homogeneous traffic flow
manifesting itself in the mentioned above phase transitions and giving
rise to the hysteresis in both of them. Besides, the jam is characterized
by the vehicle flows at different lanes being independent of one another.
We specify a certain simplified model in order to study the general
features of the car cluster self-formation under the phase transition
``free flow $\leftrightarrow$ synchronized motion''. In particular, we
show that the main local parameters of the developed cluster are
determined by the state characteristics of vehicle motion only.
\end{abstract}
]

\section{Introduction. Macroscopic models for multilane traffic dynamics}
\label{sec:intro}

The study of traffic flow formed actually a novel branch of physics since
the pioneering works by Lighthill and Whitham \cite{LW55}, Richards
\cite{R56}, and, then, by Prigogine and Herman \cite{PH71}. It is singled
out by the fact that in spite of {\it motivated}, i.e. a non-physical
individual behavior of moving vehicles (they make up a so-called ensemble
of ``self-driven particles'', see, e.g., \cite{a1,a2,a3}), traffic flow
exhibits a wide class of critical and self-organization phenomena met in
physical systems (for a review see \cite{KL1,KL2,KL3}). Besides, the
methods of statistical physics turn out to be a useful basis for the
theoretical description of traffic dynamics \cite{H97}.

The existence of a new basic phase in vehicle flow on multilane highways
called the synchronized motion was recently discovered by Kerner and
Rehborn \cite{KR2}, impacting significantly the physics of traffics as a
whole. In particular, it turns out that the spontaneous formation of
moving jams on highways proceeds mainly through a sequence of two
transitions: ``free flow $\rightarrow$ synchronized motion $\rightarrow$
stop-and-go pattern'' \cite{K98}. All the three traffic modes are phase
states, meaning their ability to persist individually for a long time.
Besides, the two transitions exhibit hysteresis \cite{K98,KR1,KR3}, i.e.,
for example, the transition from the free flow to the synchronized mode
occurs at a higher density and lower velocity than the inverse one. As
follows from the experimental data \cite{KR2,KR1,KR3} the phase transition
``free flow $\leftrightarrow$ synchronized mode'' is essentially a
multilane effect. Recently Kerner \cite{KL1,KL2,KL3} assumed it to be
caused by ``Z''-like form of the overtaking probability depending on the
car density.

The synchronized mode is characterized by substantial correlations in the
car motion along different lanes because of the lane changing maneuvers.
So, to describe such phenomena a multilane traffic theory is required.
There have been proposed several macroscopic models dealing with multilane
traffic flow and  based on the gas-kinetic theory
\cite{Ha97,HG97,HT98,HT99,HS99,Siam}, a compressible fluid model
\cite{L97} generalizing the approach by Kerner and Konh\"{a}user
\cite{Km1,Km2}, and actually a model \cite{T99a,T99b} dealing with the
time dependent Ginzburg-Landau equation.

All these models describe traffic flow in terms of the car density $\rho$,
mean velocity $v$, and, may be, the velocity variance $\theta$ or ascribe
these quantities to vehicle flow at each lane $\alpha$ individually. In
other words, the quantities $\{\rho, v,\theta\}_{\alpha}$ are  regarded as
a complete system of the traffic flow state variables and if they are
fixed then all the vehicle flow characteristics should be determined.
The given models relate the self-organization phenomena actually to the
vehicle flow instability caused by the delay in the driver response to
changes in the motion of the nearest cars ahead. In fact, let us briefly
consider their simplified version (cf. \cite{Wbook,GTF5H}) which,
nevertheless, catches the basic features taken into account:
\begin{eqnarray}
\frac{\partial \rho }{\partial t}+\frac{\partial (\rho v)}{\partial x}
&=&0\,,
\label{intro:1}\\
 \frac{\partial v}{\partial t}+v\frac{\partial
v}{\partial x} &=&-\frac{1}{ \rho }\frac{\partial {\mathcal{P}}}{\partial
x}+\frac{1}{\tau ^{\prime }}({\mathcal{U}}-v)\,.
\label{intro:2}
\end{eqnarray}
Here the former term on the right-hand side of Eq.~(\ref{intro:2}), a
so-called pressure term, reflects dispersion effects due to the finite
velocity variance $\theta $ of the vehicles and the latter one describes
the relaxation of the current velocity within the time $\tau ^{\prime }$
to a certain equilibrium value ${\mathcal{U}}\{\rho ,\theta \}$. In
particular, for
\begin{equation}
{\mathcal{P}}\{\rho ,v,\theta \}=\rho \theta -\eta \frac{\partial v}
{\partial x}\,,
\label{nolabel}
\end{equation}
where $\eta $ is a ``viscosity'' coefficient and the velocity variance
$\theta $ is treated as a constant, we are confronted with the
Kerner-Konh\"{a}user model \cite{Km1,Km2}. In the present form the
relaxation time $\tau ^{\prime }$ characterizes the acceleration
capability of the mean vehicle as well as the delay in the driver control
over the headway (see, e.g.,
\cite{hw2,HtoK,hw1}). The value of the relaxation time is typically
estimated as $\tau ^{\prime }\sim 30\,$s because it is mainly determined
by the mean time of vehicle acceleration which is a slower process than
the vehicle deceleration or the driver reaction to changes in the headway.
The equilibrium velocity ${\mathcal{U}}\{\rho \}$ (here the fixed velocity
variance $\theta $ is not directly shown in the argument list) is chosen
by drivers keeping in mind the safety, the readiness for risk, and the
legal traffic regulations. For homogeneous traffic flow the equilibrium
velocity ${\mathcal{U}}\{\rho \}=\vartheta (\rho )$ is regarded as a
certain phenomenological function meeting the conditions:
\begin{equation}
\frac{d\vartheta (\rho )}{d\rho }<0\quad \text{and}\quad \rho \vartheta
(\rho )\rightarrow 0\quad \text{as}\quad \rho \rightarrow \rho _{0}\,,
\label{intro:3}
\end{equation}
where $\rho _{0}$ is the upper limit of the vehicle density on the road.
Since the drivers anticipate risky maneuvers of distant cars also, the
dependence ${\mathcal{U}}\{\rho \}$ is nonlocal. In particular, it is
reasonable to assume that the driver behavior is mainly governed by the
car density $\rho $ at a certain distant ``interaction point''
$x_{a}=x+L^{\ast } $ rather than at the current one $x$, which gives rise
to a new term in Eq.~(\ref{intro:2}) basing on the gas-kinetic theory
\cite{HG97,HT98}. Here, for the sake of simplicity following \cite{Wbook},
we take this effect into account expanding $\rho (x+L^{\ast })$ and, then,
${\mathcal{U}}\{\rho \}$ into the Taylor series and write:
\begin{equation}
{\mathcal{U}}\{\rho \}=\vartheta (\rho )-v_{0}\frac{L^{\ast}}{\rho
}\frac{\partial \rho }{\partial x}\,,
\label{intro:4a}
\end{equation}
where
$v_{0}$ is a certain characteristic velocity of the vehicles. Then
linearizing the obtained system of equations with respect to the small
perturbations $\delta \rho ,\delta v\propto \exp (\gamma t+ikx)$ we obtain
that the long-wave instability will occur if (cf. \cite{Km1,Km2,Wbook})
\begin{equation}
\label{intro:5}
\tau ^{\prime }\left( \rho \vartheta _{\rho }^{\prime
}\right) ^{2}>v_{0}L^{\ast }+\tau ^{\prime }\theta \,,
\end{equation}
in the long-wave limit the instability increment $\mbox{Re}\,\gamma $
depends on $k$ as
\begin{equation}
\mbox{Re}\,\gamma =k^{2}\left[ \tau ^{\prime }(\rho \vartheta _{\rho
}^{\prime })^{2}-(v_{0}L^{\ast }+\tau ^{\prime }\theta )\right] \,,
\label{intro:6}
\end{equation}
and the upper boundary $k_{\text{max}}$ of the instability region in the
$k$-space is given by the expression:
$$
k_{\text{max}}^{2}=\frac{\rho }{\tau \eta }\biggl\{ \biggl[ \frac{\tau
^{\prime }(\rho \vartheta_{\rho }^{\prime })^{2}}{(v_{0}L^{\ast }+\tau
^{\prime }\theta )}\biggr] ^{1/2}-1\biggr\}
$$
(here $\vartheta _{\rho }^{\prime }=d\vartheta (\rho )/d\rho $). As
follows from (\ref{intro:5}) the instability can occur when the delay time
$\tau ^{\prime }$ exceeds a certain critical value $\tau _{c}$ and for
$\tau ^{\prime }<\tau _{c}$ the homogeneous traffic flow is stable at
least with respect to small perturbations. Moreover, the instability
increment attains its maximum at $k\sim k_{\text{max}}$, so, special
conditions are required for a wide vehicle cluster to form
\cite{Km1,Km2,Km3,Km4}. In particular, in the formal limit $\tau ^{\prime
}\rightarrow 0$ from Eq.~(\ref{intro:2}) we get
\begin{equation}
v=\vartheta (\rho )-\frac{D}{\rho }\frac{\partial \rho }{\partial x}\,,
\label{intro:7}
\end{equation}
where $D=v_{0}L^{\ast }$ plays the role of the diffusion coefficient of
vehicle perturbations. The substitution of (\ref{intro:7}) into
(\ref{intro:1}) yields the Burgers equation
\begin{equation}
\frac{\partial \rho }{\partial t}+\frac{\partial \left[ \rho \vartheta
(\rho )\right] }{\partial x}=D\frac{\partial ^{2}\rho }{\partial x^{2}}\,,
\end{equation}
which describes the vehicle flux stable, at least, with respect to small
perturbations in the vehicle density $\rho $.

However, the
recent experimental data \cite{KR2,K98,KR1,KR3,Last} about
traffic flow on German highways have demonstrated that the characteristics
of the multilane vehicle motion are more complex (for a review see
\cite{KL1,KL2,KL3}). In particular, there are actually three types of
synchronized mode, the totally homogeneous state, homogeneous-in-speed and
totally heterogeneous ones \cite{KR2}. Especially the homogeneous-in-speed
state demonstrates the fact that, in contrast to the free flow, there is
no direct relationship between the density and flux of vehicles in the
synchronized mode, because their variations are totally uncorrelated
\cite{KR2}. For example, an increase in the vehicle density can be
accompanied by either an increase or decrease in the vehicle flux, with
the car velocity being practically constant. As a result, the synchronized
mode corresponds to a two-dimensional region in the flow-density plane
($j\rho$-plane) rather than to a certain line $j=\vartheta(\rho)\rho$
\cite{KR2}. Keeping in mind a hypothesis by Kerner \cite{KL2,KL3,KL2a}
about the metastability of each particular state in this synchronized mode
region it is natural to assume that there should be at least one
additional state variable affecting the vehicle flux. The other important
feature of the synchronized mode is the key role of some cars bunched
together and traveling much faster than the typical ones, which enables us
to regard them as a special car group \cite{KR2}. Therefore, in the
synchronized mode the function of car distribution in the velocity space
should have two maxima and we will call such fast car groups platoons in
speed.

Anomalous properties of the synchronized mode have been substantiated also
in \cite{Last} using single-car-data. In particular, as the car density
comes to the critical value $\rho_c$ of the transition ``free flow
$\leftrightarrow$ synchronized mode'' the time-headway distribution
exhibits a short time peak (at 0.8~sec.). This short time-headway
corresponds to
``...platoons of some vehicles traveling very fast--their drivers are
taking the risk of driving ``bumper-to-bumper'' with a rather high speed.
These platoons are the reason for occurrence of high-flow states in free
traffic'' \cite{Last}. The platoons are metastable and their destruction
gives rise to the congested vehicle motion \cite{Kplat}. In the
synchronized mode the weight of the short time-headways is less, however,
almost every fourth driver falls below the 1-sec-threshold. In the
vicinity of the transition ``free flow $\leftrightarrow$ synchronized
mode'' the short time-headways have the greatest weight. In other words,
at least near the given phase transition the traffic flow state is to be
characterized by two different driver groups which separate from each
other in the {\it velocity} space and, consequently, in multilane traffic
flow there should be another relaxation process distinct from one taken
into account by the model~(\ref{intro:1}), (\ref{intro:2}). In order to
move faster than the statistically averaged car a driver should
permanently manoeuvring pass by the cars moving ahead. Meeting of several
such ``fast'' drivers seems to cause the platoon formation. Obviously, to
drive in such a manner requires additional efforts, so, each driver needs
a certain time $\tau$ to get the decision whether or not to take part in
these manoeuvres. Exactly the time $\tau$ characterizes the relaxation
processes in the platoon evolution. It should be noted that the overtaking
manoeuvres are not caused by the control over the headway distance and,
thus, the corresponding  transient processes may be much slower then the
driver response to variations in the headway to prevent possible traffic
accidents.

The analysis of the obtained optimal-velocity function $V(\Delta x)$
demonstrates its dependence not only on the headway $\Delta x$ but also on
the local car density. So, in congested flow the drivers supervises the
vehicle arrangement or, at least, try to do this in a sufficiently large
neighborhood covering several lanes.

Another unexpected fact is that the synchronized mode is mainly
distinctive not due to the car velocities at different lanes being equal.
In the observed traffic flow various lanes did not exhibit a substantial
difference in the car velocity even in the free flow. In agreement with
the results obtained by Kerner \cite{KR2} the synchronized mode is singled
out by small correlations between fluctuations in the car flow, velocity
and density. There is only a strong correlation between the velocities at
different lanes taken at the same time, however, it decreases sufficiently
fast as the time difference increases. By contrast, there are strong
long-time correlations between the flow and density for the free flow  as
well as the stop-and-go mode. In these phases the vehicle flow directly
depends on the density.

Thereby, the free flow, the
synchronized mode and the jammed motion seem
to be qualitatively distinct from one another at the microscopic level.
So, it is likely that to describe macroscopically traffic phase
transitions the set of the state variables $\{\rho, v,\theta\}_{\alpha}$
should be completed with an additional pa\-ra\-me\-ter (or parameters)
reflecting the {\it internal} correlations in the car dynamics. In other
words, this parameter has to be due to the ``many-body'' effects in the
car interaction in contrast to such {\it external} variables as the mean
car density and velocity being actually the zeroth and first moments of
the ``one-particle'' distribution function. Thus, it can be regarded as an
independent state variable of traffic flow. The derivation of macroscopic
traffic equations based on a Boltzmann like kinetic approach  \cite{W98}
has also shown that there is an additional internal degrees of freedom in
the vehicle dynamics.

\begin{figure} \centerline{\psfig{file=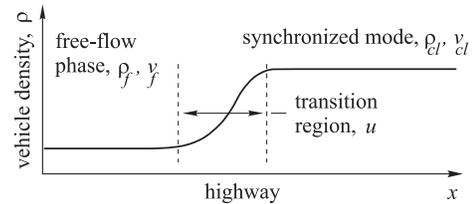,width=63mm}}
\caption{Transition region separating, e.g., the free-flow and
synchronized mode.} \label{F1}
\end{figure}

In any case a theory of unstable traffic flow has to answer, in
particular, to a question of why its two phases, e.g., the free flow and
the synchronized mode, can coexist and, thus, what is the difference
between them as well as why the separating  transition region
(Fig.~\ref{F1}) does not widen but keeps a certain thickness. Besides, it
should specify the velocity $u$ of this region depending on the traffic
phase characteristics. There is a general expression relating the
transition region velocity $u$ to the density and mean velocity of cars in
the free flow and a developed car cluster: $\rho _{f}$, $v_{f}$, and $\rho
_{cl}$, $v_{cl}$, respectively, that follows from the vehicle conservation
\cite{LW55}, namely, the Lighthill--Whitham formula:
\begin{equation}
u=\frac{\rho _{cl}v_{cl}-\rho _{f}v_{f}}{\rho _{cl}-\rho _{f}}\,.
\label{intro:4}
\end{equation}
A specific model is to give additional relationships between the
quantities $u$, $\rho _{f}$, $v_{f}$, and $\rho _{cl}$, $v_{cl}$ resulting
from particular details of the car interaction. We note that a description
similar to Eqs.~(\ref{intro:1}), (\ref{intro:2}) dealing solely with the
external parameters $\{\rho,v\}$ do not actually make a distinction
between the free-flow and congested phases and their coexistence is due to
the particular details of the car interaction.

The transition ``free flow $\leftrightarrow$ synchronized motion'' is rather
similar to aggregation phenomena in physical systems such as
undercooled liquid when in a metastable phase (undercooled liquid) the
transition to a new ordered (crystalline) phase goes through the formation of
small clusters. Keeping in mind this analogy Mahnke {\it et al.}
\cite{MP97,MK99,KM00} have proposed a kinetic approach
based on stochastic master equation
describing the
synchronized mode formation that deals with individual free cars and their
clusters. The cluster evolution is governed by the attachment and
evaporation of the individual cars and the synchronized mode is regarded as
the motion of a large cluster.

To describe such phenomena in physical systems it was developed an effective
macroscopic approach, called the Landau phase transition theory
\cite{Landau}, that introduces a certain order parameter $h$ characterizing the
correlations, e.g., in the atom arrangement. In the present paper
following practically the spirit of the Landau theory we develop a
phenomenological approach to the description of the traffic flow instability
that ascribes to the vehicle flux an additional {\it internal} parameter
which will be also called the order parameter $h$ and allows for the effect of
lane changing on the vehicle motion. In this way the free flow and the congested
phases become in fact distinctive and solely the conditions of their coexistence
as well the dynamics of the transition layer are the subject of specific models.

\section{Order parameter and the individual driver behavior}\label{sec:2}

We describe the vehicle flow on a multilane highway in terms of its
characteristics averaged over the road cross-section, namely, by the car
density $\rho $, the mean velocity $v$, and the order parameter $h$. The
latter is the measure of the correlations in the car motion or, what is
equivalent, of the car arrangement regularity forming due to the lane
change by the ``fast'' drivers. Let us discuss the physical meaning of the
order parameter $h$ in detail considering the free flow, synchronized mode
and jammed traffic individually (Fig.~\ref{FHDef}).

\subsection{Physical meaning of the order parameter $h$ and its governing
equation}

\begin{figure}
\centerline{\psfig{file=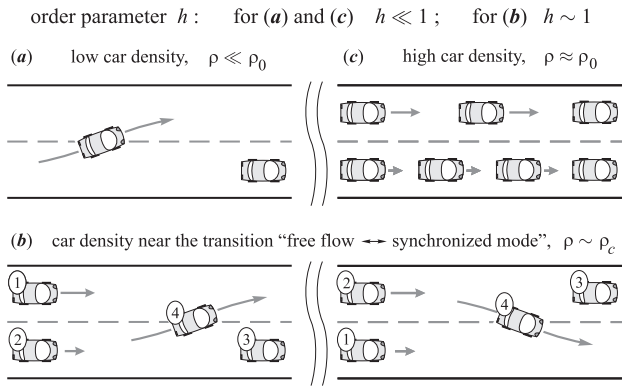,width=82mm}} \caption{Schematic
illustration of the car arrangement in the various phases of traffic flow
and the multilane vehicle interaction caused by car overtaking.}
\label{FHDef}
\end{figure}

When vehicles move on a multilane highway without changing the lanes they
interact practically with the nearest neighbors ahead only and, so, there
should be no internal correlations in the vehicle flow at different lanes.
Therefore, although under this condition the traffic flow can exhibit
complex behaviour, for example, the ``stop-and-go'' waves can develop, it
is actually of one-dimensional nature. In particular, the drivers that
would prefer to move faster than then the statistically mean driver will
bunch up forming the platoons headed by a relatively slower vehicle. When
the cars begin to change lanes for overtaking slow vehicles the car
ensembles at different lanes will affect one another. The case of this
interaction is due to that a car during a lane change manoeuvre occupies,
in a certain sense, two lanes simultaneously, affecting the cars moving
behind it at both the lanes. Figure~\ref{FHDef}$b$ illustrates this
interaction for cars 1 and 2 through car 4 changing the lanes. The drivers
of both cars 1 and 2 have to regard car 4 as the nearest neighbor and, so,
their motion will be correlated during the given manoeuvre and after it
during the relaxation time $\tau'$. In the same way car 1 is affected by
car 3 because the motion of car 4 directly depends on the behavior of car
3. The more frequently lane changing is performed, the more correlated
traffic flow on a multilane highway. Therefore, it is reasonable to
introduce the order parameter $h$ being the mean density of such car
triplets normalized to its maximum possible for the given highway and to
regard it as a measure of the mulitlane correlations in the vehicle flow.

On the other hand the order parameter $h$ introduced in this way can be
regarded as a measure of the vehicle arrangement regularity. Let us
consider this question in detail for the free flow, synchronized mode, and
jammed traffic individually. In the free flow the feasibility of
overtaking makes the vehicle arrangement more regular because of platoon
dissipation. So as the order parameter $h$ grows the free traffic becomes
more regular. Nevertheless, in this case the density of the car mulitlane
triplets remains relatively low, $h\ll 1$, and the vehicle ensembles
should exhibit weak correlations. Whence it follows also that the mean car
velocity $\vartheta $ is an increasing function of the order parameter $h$
in the free flow. In the jammed motion (Fig.~\ref{FHDef}$c$) leaving
current lanes is hampered because of lack of room for the manoeuvres. So
the car ensembles at different lanes can be mutually independent in spite
of individual complex behavior. In the given case the order parameter must
be small too, $h\ll 1$, but, in contrast, the car mean velocity should be
a decreasing function of $h $. In fact, for highly dense traffic any lane
change of a car requires practically that the neighbor drivers
decelerating give place for this manoeuvres.

\begin{figure}
\centerline{\psfig{file=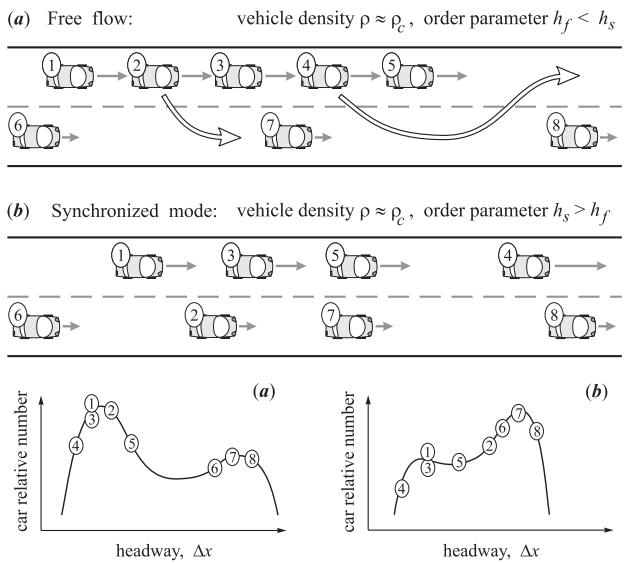,width=82mm}}\caption{Schematic
illustration of the alteration in the vehicle arrangement near the phase
transition ``free flow $\leftrightarrow$ synchronized mode''.}
\label{FHPhT}
\end{figure}

Figure~\ref{FHPhT} illustrates the transition ``free flow $\leftrightarrow
$ synchronized mode''. As the car density grows in free flow, the ``fast''
drivers that at first overtake slow vehicles individually begin to gather
into platoons headed by more ``slow'' cars among them but, nevertheless,
moving faster than the statistically mean vehicle\ (Fig.~\ref{FHPhT}$a$).
The platoons are formed by drivers preferring to move as fast as possible
keeping short headways without lane changing. Such a state of the traffic
flow should be sufficiently inhomogeneous and the vehicle headway
distribution has to contain a short headway\ spike as observed
experimentally in \cite{Last}. Therefore, even at a sufficiently high car
density the free flow  should be characterized by weak multilane
correlations and not too great values of the order parameter $h_{f}$. The
structure of these platoons is also inhomogeneous, they comprise cars
whose drivers would prefer to move at different headways (for a fixed
velocity) under comfortable conditions, i.e., when the cars moving behind
a given car do not jam it or none of the vehicles moving on the
neighboring lanes hinders its motion at the given velocity provided it
changes the current lane. So, when the density of vehicles attains
sufficiently high values and their mean velocity decreases remarkably with
respect to the velocity on the empty highway some of the ``fast'' drivers
can decide that there is no reason to move so slowly at such short
headways requiring strain. Then they can either overtake the car heading
the current platoon by changing lanes individually or leave the platoon
and take vacant places (Fig.~\ref{FHPhT}$a$). The former has to increase
the multilane correlations and, in part, to decrease the mean vehicle
velocity because the other drivers should give place for this manoeuvres
in a sufficiently dense traffic flow. The latter also will decrease the
mean vehicle velocity because these places were vacant from the standpoint
of sufficiently ``fast'' drivers only but not from the point of view of
the statistically mean ones preferring to keep longer headways in
comparison with the platoon headways. Therefore, the statistically mean
drivers have to decelerate, decreasing the mean vehicle velocity. The two
manoeuvre types make the traffic flow more homogeneous dissipating the
platoons and smoothing the headway distribution (Fig.~\ref {FHPhT}$b$ and
the low fragment). Besides, the single-vehicle experimental data
\cite{Last} show that the synchronized mode is singled out by long-distant
correlations in the vehicle velocities, whereas the headway fluctuations
are correlated only on small scales, which justifies the assumptions of
the synchronized mode being a more homogeneous state than the free flow.
We think that the given scenario describes the synchronized mode formation
which must be characterized by a great value of the order parameter,
$h_{s}>h_{f}$, and a lower velocity in comparison with the free flow at
the same vehicle density.

In addition, whence it follows that, first, the left boundary of the
headway distribution should be approximately the same for both the free
flow and the synchronized mode near the phase transition, which
corresponds the experimental data \cite{Last}. Second, since in this case
the transition from the free flow to the synchronized mode leads to the
decrease in the mean velocity, the ``fast'' driver will see no reason to
alter their behaviour and to move forming platoons again until the vehicle
density decreases and the mean velocity grows enough. It is reasonable to
relate this characteristics to the experimentally observed hysteresis in
the transition ``free flow $\leftrightarrow $ synchronized mode''
\cite{K98,KR1,KR3}. Third, for a car to be able to leave a given platoon
the local vehicle arrangement at the neighboring lane should be of special
form and when an event of the vehicle rearrangement occurs its following
evolution depends also on the particular details of the neighboring car
configuration exhibiting substantial fluctuations. Therefore, the
synchronized mode can comprise a great amount of local metastable states
and
correspond to a certain two-dimensional region on the flow-density plane
($j\rho $-plane) rather than a line $j=\vartheta (\rho )\rho$, which
matches the experimental data \cite{KR2} and the modern notion of the
synchronized mode nature \cite {KL1,KL2,KL3}. This feature seems to be
similar to that met in physical media with local order, for example, in
glasses where phase transitions are characterized by wide range of
controlling parameters (temperature, pressure, {\it etc.}) rather than
their fixed values (see, e.g., \cite{Ziman}).

This uncertainty of the synchronized mode, at least qualitatively, may be
regarded as an effect of the internal fluctuations of the order parameter
$h$ and at the first step we will ignore them assuming the order parameter
$h$ to be determined in a unique fashion for fixed values of the vehicle
density $\ \rho $ and the mean velocity $v$. Thus for a uniform vehicle
flow we write:
\begin{equation}
\tau \frac{dh}{dt}=-\Phi (h,\rho ,v)\,,
\label{2.1}
\end{equation}
where $\tau $ is the time required of drivers coming to the decision to
begin or stop overtaking manoeuvres and the function $\Phi(h,\rho ,v)$
possesses a single stationary point $h=h(\rho,v)$ being stable and, thus,
\begin{equation}
\frac{\partial \Phi }{\partial h}>0\,.
\label{2.2}
\end{equation}
The latter inequality is assumed to hold for all the values of the order
parameter for simplicity. We note that equation~(\ref{2.1}) also allows
for the delay in the driver response to changes on the road. However, in
contrast with models similar to (\ref{intro:1}) and (\ref{intro:2}), here
this effect is not the origin of the traffic flow instability and, thus,
its particular description is not so crucial. Moreover, as discussed in
the Introduction, the time $\tau$ characterizes the delay in the driver
decision concerning to the lane changing but not the control over the
headway, enabling us to assume $\tau\gg\tau'$.

\begin{figure}[h]
\centerline{\psfig{file=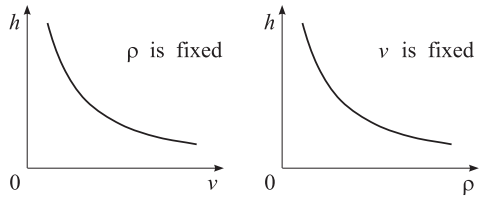,width=63mm}}
\caption{Qualitative sketches of the order parameter
$h$ as a function of the vehicle mean velocity $v$ and the
density $\rho$ specified by the behavior of individual drivers.
}
\label{F4}
\end{figure}

The particular value $h(v,\rho )$ of the order parameter results from the
compromise between the danger of accident during changing lanes and the will
to move as fast as possible. Obviously,
the lower is the mean vehicle ve\-lo\-ci\-ty $v$ for a fixed value of $\rho $,
the weaker is the lane changing danger and the stronger is the will to move
faster. Besides, the higher is the vehicle density $\rho $ for a fixed value
of $v$, the stronger is this danger (here the will has no effect at all).
These statements enable us to regard the dependence $h(v,\rho )$ as
a decreasing function of both the variables $v$, $\rho $
(Fig.~\ref{F4}) and taking into account inequality~(\ref{2.2}) to write:
\begin{equation}
\frac{\partial \Phi }{\partial v}>0\,,\quad \frac{\partial \Phi }{\partial
\rho }>0\,,
\label{2.3}
\end{equation}
with the latter inequality stemming from the danger effect only.

\begin{figure}
\centerline{\psfig{file=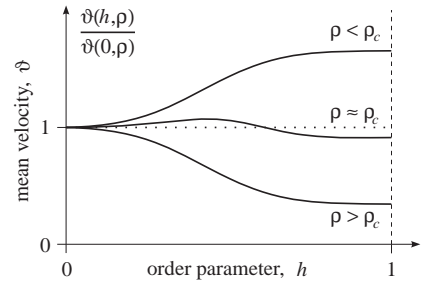,width=55mm}}
\caption{A qualitative sketch of the mean vehicle
velocity vs. the order parameter $h$ for several fixed values of the
vehicle density $\rho$.
} \label{F5}
\end{figure}

Equation (\ref{2.1}) describes actually the behavior of the drivers who
prefer to move faster than the statistically mean vehicle and whose
readiness for risk is greatest. Exactly this group of drivers govern the
value of $h$. There is, however, another characteristics of the driver
behavior, it is the mean velocity $v=\vartheta (h,\rho )$ chosen by the
{\it statistically averaged} driver taking into account also the danger
resulting from the frequent lane changing by the ``fast'' drivers. This
characteristics is actually the same as one discussed in the Introduction
but depends also on the order parameter $h$. So, as a function of $\rho $
it meets conditions~(\ref{intro:3}). Concerning the dependence of
$\vartheta(h,\rho)$ on $h$ we can state that generally this function
should be increasing for small values of the car density, $\rho \ll \rho
_{0}$, because in the given  case the lane changing practically makes no
danger to traffic and all the drives can overtake vehicle moving at lower
speed without risk. By contrast, when the vehicle density is sufficiently
high, $\rho \lesssim \rho _{0}$, only the most ``impatient'' drivers
permanently change the lanes for overtaking, making an additional danger
to the most part of other drivers. Therefore, in this case the velocity
$\vartheta (h,\rho )$ has to decrease as the order parameter $h$
increases. For certain intermediate values of the vehicle density,
$\rho\approx\rho_{c}$, this dependence is to be weak. Fig.~\ref{F5} shows
the velocity $\vartheta(h,\rho)$ as a function of $h$ for different values
of $\rho$, where, in addition, we assume the effect of order parameter
$h\in (0,1)$ near the boundary points weak and set
\begin{equation}
\frac{\partial \vartheta }{\partial h}=0\quad \text{at}\quad h=0\
\text{and}\ h=1\,. \label{2.3a}
\end{equation}

We will ignore the delay in the relaxation of the mean velocity to the
equilibrium value $v=\vartheta (h,\rho)$ because the corresponding delay
time characterizes the driver control over the headway and should be
short, as already discussed above. Then the governing equation~(\ref{2.1})
for the order parameter $h$ can be rewritten in the form:
\begin{equation}
\tau \frac{dh}{dt}=-\phi (h,\rho )\,;\quad
\phi(h,\rho )\stackrel{\text{def}}{=}\Phi [h,\rho ,\vartheta (h,\rho )]\,.
\label{2.4}
\end{equation}
For the steady state uniform vehicle flow the solution of the equation $\phi
(h,\rho )=0$ specifies the dependence $h(\rho )$ of the order parameter on
the car density. Let us, now, study its properties and stability.

\subsection{Nonmonotony of the $h(\rho )$ dependence and the traffic flow
instability}

To study the local characteristics of the right-hand side of Eq.~(\ref{2.4})
we analyze its partial derivatives
\begin{eqnarray}
\frac{\partial \phi }{\partial h} &=&\frac{\partial \Phi }{\partial h}+
\frac{\partial \Phi }{\partial v}\frac{\partial \vartheta }{\partial h}\,,
\label{2.6} \\
\frac{\partial \phi }{\partial \rho } &=&\frac{\partial \Phi }{\partial \rho}
+\frac{\partial \Phi }{\partial v}\frac{\partial \vartheta }{\partial \rho }
\,.  \label{2.7}
\end{eqnarray}
As mentioned above, the value of $\partial \Phi /\partial \rho $ is solely
due to the danger during changing lanes, so this term can be ignored until
the vehicle density $\rho$ becomes sufficiently high. In other words, in a
certain region $\rho <\rho_h\lesssim \rho _{0}$ the derivative $\partial \phi
/\partial \rho \sim (\partial \Phi /\partial v)(\partial \vartheta /\partial
\rho )<0$ by virtue of (\ref{intro:3}) and (\ref{2.3}). So, the local
behavior of the function $h(\rho)$ (meeting the equality $d\phi=0$ and, thus,
$dh/d\rho = -(\partial\phi/\partial\rho)(\partial\phi/\partial h)^{-1}$)
depends directly on the sign of the derivative $\partial \phi /\partial h$,
it is increasing or decreasing  for $\partial \phi /\partial h>0$ or
$\partial \phi /\partial h<0$, respectively.

For long-wave perturbations $\propto \exp \{ikx\}$ of the car
distribution on a highway the density $\rho $ can be treated as a constant
at the lower order in $k$. Therefore, according to Eq.~(\ref{2.4})
the steady state traffic flow is unstable if $\partial \phi /\partial h<0$.

\begin{figure}
\centerline{\psfig{file=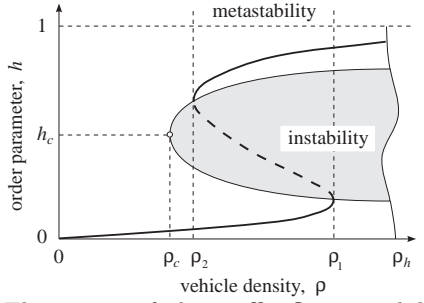,width=55mm}} \caption{The region of the
traffic flow instability in the $h\rho$-plane and the form of the curve
$h(\rho)$ displaying the dependence of the order parameter on the vehicle
density.
The plot is a qualitative sketch.}
\label{F6}
\end{figure}

Due to (\ref{2.2}) and (\ref{2.3a}) the first term on the right-hand side of
(\ref{2.6}) is dominant in the vicinity of the lines $h=0$ and $h=1$, thus,
in this region the curve $h(\rho )$ is increasing and the stationary state
of the traffic flow is stable. For $\rho <\rho _{c}$ the value $\partial
\vartheta /\partial h >0$ (Fig.~\ref{F5}), therefore, the whole region
$\left\{ 0<h<1,\;0<\rho <\rho _{c}\right\} $ corresponds to the stable
car motion. However, for $\rho >\rho _{c}$ there can be a region of the
order parameter $h$ where the derivative $\partial \phi /\partial h$ changes
the sign and the vehicle motion becomes unstable. Indeed, the solution
$v=\eta (h,\rho )$ of the equation $\Phi (h,\rho ,v)=0$ can be regarded as
the mean vehicle velocity controlled by the ``fast'' drivers and is
decreasing function of $h$ because of $\partial \eta /\partial h=-(\partial
\Phi /\partial h)/(\partial \Phi /\partial v)^{-1}$. So, once such ``active''
drivers become to change the lanes to move faster, they will do this as
frequently as possible especially if the mean velocity decreases, which
corresponds to a considerable increase in $h$ for a small decrease in $v$.
So, it is quite natural to assume that the value of $\partial \eta /\partial
h$ for $\rho >\rho _{c}$ is sufficiently small and
\begin{equation}
\frac{\partial \phi }{\partial h} =
\frac{\partial\Phi}{\partial v}\left(
\frac{\partial \vartheta }{\partial h}-\frac{\partial \eta }{\partial h}
\right)<0\,.
\end{equation}
Under these conditions the instability region does exist, the curve $h(\rho)$
can look like ``S''
(Fig.~\ref{F6}) and its decreasing branch corresponds to the unstable
vehicle flow. The lower increasing branch  matches the free flow state of
the car motion, whereas the upper one should be related to the
synchronized mode because it is characterized by the order parameter
coming to unity.

\subsection{Hysteresis and the fundamental diagram}

The obtained dependence $h(\rho)$ actually describes the first order phase
transition in the vehicle motion. Indeed, when increasing the car density
exceeds the value $\rho_1$ the free flow becomes absolutely unstable and
the synchronized mode forms through a sharp jump of the order parameter. If,
however, after that the car density decreases  the synchronized mode will persist
until the car density attains the value $\rho_2 < \rho_1$. It is a typical
hysteresis and the region $(\rho_2, \rho_1)$ corresponds to the metastable
phases of traffic flow.

\begin{figure}
\centerline{\psfig{file=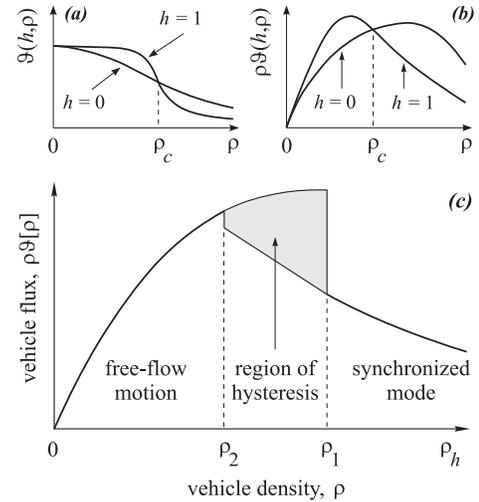,width=63mm}} \caption{The mean vehicle
velocity ({$a$}) and the vehicle flux ({$b$}) vs. the vehicle density for
the limit values of the order parameter $h=0$ and $h=1$ as well as the
resulting fundamental diagram ({$c$}). A qualitative sketch.} \label{F7}
\end{figure}

Let us, now, discuss a possible form of the fundamental diagram
$j=j(\rho)$
showing the vehicle flux $j=\rho \vartheta [\rho ]$ as a function of the car
density $\rho $, where, by definition, $\vartheta [\rho ]=\vartheta
[h(\rho ),\rho ]$. It should be pointed out that here we confine our
consideration to the region of not too large values of the car density,
$\rho<\rho_h$, where the transition ``free flow $\leftrightarrow$
synchronized mode'' takes place. The transition ``synchronized mode
$\leftrightarrow$ jammed traffic'' will be discussed below.
Fig.~\ref{F7}{$a$} displays the dependence $\vartheta (h,\rho )$ of the
mean vehicle velocity on the density $\rho $ for the fixed limit values of
the order parameter $h=0$ and 1. For a small values of $\rho $ these
curves practically coincide with each other. As the vehicle density $\rho
$ grows and until it comes close to the critical value $\rho _{c}$ when
the lane change danger becomes substantial, the velocity $\vartheta
(1,\rho )$ practically does not depend on $\rho $. So at the point $\rho
_{c}$ at which the curves $\vartheta (1,\rho )$ and $\vartheta (0,\rho )$
meet each other, $\vartheta (1,\rho )$ is to exhibit sufficiently sharp
decrease in comparison with the latter one. Therefore, on one hand, the
function $j_{1}(\rho )=\rho \vartheta (1,\rho )$ has to be decreasing for
$\rho > \rho _{c}$. On the other hand, at the point $\rho _{c}$ for $h\ll
1$ the effect of the lane change danger is not extremely strong, it only
makes the lane change ineffective, $\partial \vartheta /\partial h\approx
0$ (Fig.~\ref{F5}). So it is reasonable to assume the function $j_{0}(\rho
)=\rho \vartheta (0,\rho )$ increasing near the point $\rho _{c}$. Under
the adopted assumptions the relative arrangement of the curves $j_{0}(\rho
)$, $j_{1}(\rho )$ is demonstrated in Fig.~\ref{F7}{$b$}, and
Fig.~\ref{F7}{$c$} shows the fundamental diagram of traffic flow resulting
from Fig.~\ref{F6} and Fig.~\ref{F7}{$b$}.

Concluding the present section we note that in the given description of the
driver behavior governing the order parameter $h$ the vehicle flux $j(h,\rho
)=\rho \vartheta (h,\rho )$ is an external characteristics of traffic flow.
So, the obtained form of the fundamental diagram does not follows directly
from the developed model,
but can be interpreted sufficiently reasonable.
It can be rigorously justified if the critical point $\rho_{c}$ corresponds
to the maximum of the flux $j(h^{*},\rho )$ for a certain fixed value $h^{*}$
of the order parameter. In other words, when the road capacity is exhausted
and the following increase in the vehicle density leads to a decrease in the
vehicle flux the drivers divide into two groups, the majority prefer to move
at their own lanes whereas the most ``impatient''
drivers change the lanes as
frequently as possible, giving rise to the traffic instability. This
problem, however, deserves an individual investigation.

\section{Phase coexistence. Diffusion limited cluster motion}
\label{sec:3}

The previous section has considered uniform traffic flow, so, analyzed
actually the individual characteristics of the free flow and the
synchronized mode. In the present section we study their coexistence,
i.e., the conditions under which a car cluster of finite size forms. This
problem, however, requires that the traffic flow model be defined
concretely. Therefore, in what follows we will consider a certain simple
model which illustrates the characteristic features of the car cluster
self-organization without complex mathematical manipulations.

\begin{figure}[h]
\centerline{\psfig{file=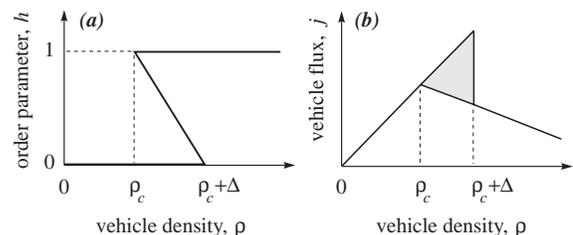}} \caption{The dependence $h(\rho)$ and
the fundamental diagram of traffic flow described by the
model~(\protect\ref{3.1}), (\protect\ref{3.2}).}
\label{F8}
\end{figure}

As before, the model under consideration assumes the mean velocity
relaxation to be immediate and modifies the governing equation~(\ref{2.4})
in such a way as to ascribe the order parameter $h$ to a local car group.
In other words, we describe the vehicle flow by the Lighthill--Whitham
equation with dissipation (see, e.g., \cite{N96} and also Introduction),
replace the time derivative in Eq.~(\ref{2.4}) by the particle derivative,
and take into account that the order parameter cannot exhibit substantial
variations over scales $l\sim \theta^{1/2} \tau \lesssim v_{0}\tau $
($\theta $ is the velocity variance, $v_0$ is the typical car velocity in
the free flow). Namely, we write:
\begin{eqnarray}
\frac{\partial \rho }{\partial t}+\frac{\partial \left[ \rho \vartheta
(h,\rho )\right] }{\partial x} &=&D\frac{\partial ^{2}\rho } {\partial x^{2}}
\,,  \label{3.1} \\
\tau \left[ \frac{\partial h}{\partial t}+\vartheta (h,\rho )\frac{\partial
h}{\partial x} \right] &=&{\hat{{\cal L}}}\{h\}-\phi (h,\rho )+\xi (x,t)\,.
\label{3.2}
\end{eqnarray}
Let us discuss the meaning of the particular terms of the given model. The
Burgers equation~(\ref{3.1}), as
already discussed in Introduction, allows for the
fact that drivers govern their motion taking into account not only the
behavior of the nearest cars, but the state of traffic flow inside the
whole field of their front view of length. The
effective diffusivity
$D$ can be estimated as $D\sim L^{*}v_{0}$, where $L^*\gg l$ is a front
distance looked through by drivers assumed to be much greater that the
scale $l$, so
\begin{equation}
D\tau \sim lL^{*}\gg l^{2}\,.
\label{3.3}
\end{equation}
The function $\phi (h,\rho )$ is of the form
\begin{equation}
\phi (h,\rho )\stackrel{\text{def}}{=}h(1-h)[a(\rho )-h]\,,
\label{3.202}
\end{equation}
where
\[
a(\rho )=\left\{
\begin{array}{lll}
1 & \;\text{for} & \rho <\rho _{c}\,, \\
(\rho _{c}+\Delta -\rho )/\Delta & \;\text{for} & \rho _{c}<\rho <\rho
_{c}+\Delta \,, \\
0 & \;\text{for} & \rho >\rho _{c}+\Delta \,.
\end{array}
\right.
\]
It describes such a driver behavior that $h=0$ and $h=1$ are the unique
stable values of the order parameter for $\rho <\rho _{c}$ and $\rho >\rho
_{c}+\Delta $, respectively, whereas, for $\rho _{c}<\rho <\rho _{c}+\Delta $
the points $h=0$, $h=1$ are both locally stable and there is an additional
unstable stationary point, namely, $h=a(\rho )$. The term
\begin{equation}
\hat{{\cal L}}\{h\}\stackrel{\text{def}}{=}
l^{2}\frac{\partial ^{2}h}{\partial x^{2}}+\frac{l}
{\sqrt{2}}\frac{\partial h}{\partial x}
\label{3.201}
\end{equation}
governs spatial variations in the field $h(x,t)$ and takes into account that
drivers mainly follow the behavior of cars in
front of them and cars moving at
the rear cannot essentially affect them. The mean car velocity depends
on $h$ and $\rho $ as
\begin{equation}
\rho \vartheta (h,\rho )\stackrel{\text{def}}{=}
\rho \vartheta _{0}(1-h)+[\rho _{c}\vartheta
_{0}-\nu (\rho -\rho _{c})]h\,.
\label{3.2b}
\end{equation}
The last term on the right-hand side of Eq.~(\ref{3.2}) characterizes
the random fluctuations in the order parameter dynamics:
\begin{eqnarray}
\label{3.203a}
\left\langle \xi (x,t)\right\rangle &=&0\,,\\
\label{3.203}
\left\langle \xi (x,t)\xi
(x^{\prime },t^{\prime })\right\rangle &=&\sigma ^{2}l\tau \delta
(x-x^{\prime})\delta (t-t^{\prime })\,,
\end{eqnarray}
where $\sigma $ is their dimensionless amplitude. Expressions
(\ref{3.202}) and (\ref{3.2b}) gives the $h(\rho )$-dependence and the
fundamental diagram shown in Fig.~\ref{F8} simplifying
the one presented in Fig~\ref{F7}.

If we ignore the random fluctuations of the order parameter $h$, i.e., set
$\sigma =0$, then the model~(\ref{3.1}), (\ref{3.2}) will give us an artificially
long delay (much greater than $\tau $) in the order parameter variations
from, for example, the unstable point $h=0$ to the stable point $h=1$. Such
a delay can lead to a meaningless great increase of the vehicle density in
the free flow  without phase transition to congestion. In order to
avoid this artifact and to allow for the effect of real fluctuations in the
driver behavior we also will assume the amplitude $\sigma $ to obey the
condition \cite{LM00}:
\begin{equation}
\left( \frac{l}{L^{*}}\right) ^{5/4}\lesssim \sigma \ll 1
\label{3.5}
\end{equation}
($\sigma \ll 1$, because, otherwise, the traffic flow dynamics would be
totally random). It should be noted that small random variations of the
order parameter $h$ near the points $h=0$, $h=1$ going into the regions $h<0$ and
$h>1$, respectively, do not come into conflict with its physical meaning
as the measure of the car motion correlations. Indeed,
the chosen values $h=0$ and $h=1$ can describe a renormalization of real
correlation coefficients $\tilde{h}=\tilde{h}_{1} > 0$, $\tilde{h}_{2} < 1$.

According to Eq.~(\ref{3.2}), for the order parameter $h$ the characteristic
scale of its spatial variations is $l$, so, the layer $\Im _{h}$ separating
the regions where $h\approx 0$ and 1 is of thickness about $l$. Due to
inequality~(\ref{3.3}) the car density on such scales can be treated as
constant. Therefore, the transition region ${\cal L}_{\rho }$ between
practically the uniform free flow and the congested phase is of thickness
determined mainly by spatial variations of the vehicle density and on such
scales the layer $\Im _{h}$ can be treated as an infinitely thin interface.
In addition, the characteristic time scale of the layer $\Im _{h}$ formation
is about $\tau $, whereas it takes about the time $\tau _{\rho }\sim
D/v_{0}^{2}\sim \tau (L^{*}/l)\gg \tau $ for the layer ${\cal L}_{\rho }$ to
form. Thereby, when analyzing the motion of a wide car clusters we may
regard the order parameter distribution $h(x,t)$ as quasi-stationary for a
fixed value of the car density $\rho $.

Let us, now, consider two possible limits of the layer $\Im _{h}$ motion
under such conditions.

\subsection{Regular dynamics}

\begin{figure}
\centerline{\psfig{file=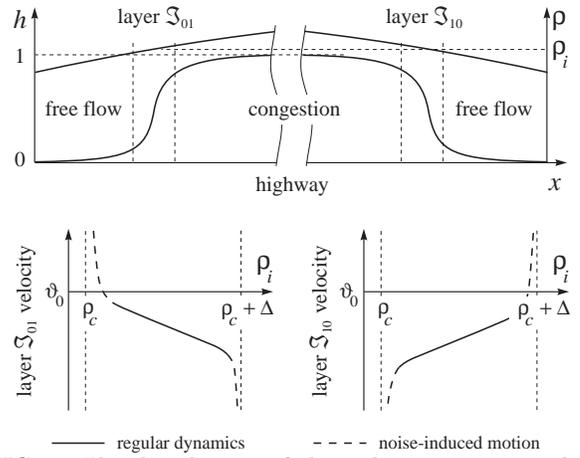}} \caption{The distribution of the order
parameter and the car density in the vicinity of the layers $\Im_h$ of the
transition between the free flow and the synchronized phase as well as the
velocity of their motion vs. the local values $\rho_i$ of the car
density.}
\label{F9}
\end{figure}

In the region $\rho _{c}<\rho <\rho _{c}+\Delta$ until the value of $a(\rho)$
comes close to the boundaries $h=0$ and $h=1$ the effect of the random
fluctuations is ignorable. In this case by virtue of the adopted
assumptions the solution of Eq.~(\ref{3.2}) that describes the layer $\Im
_{h}$ moving at the speed $u$ is of the form:
\begin{equation}
h=\frac{1}{2}\left[ 1+\tanh \left( \frac{x-ut}{\lambda }\right) \right]\,.
\label{3.4}
\end{equation}
Here for the layer $\Im _{01}$ of the transition ``free-flow $\rightarrow$
synchronized mode'' and for the layer $\Im _{10}$ of the opposite transition
(Fig.~\ref{F9})
\begin{equation}
\lambda _{01}=\frac{2\sqrt{2}}{\eta _{v}}l\,,\quad \lambda _{10}= -2\sqrt{2}
\eta _{v}l  \label{3.6}
\end{equation}
\begin{eqnarray}
u_{01} &=&\vartheta_0 - \frac{\Delta_v}{2} -\frac{l}{\sqrt{2}\eta _{v}\tau }
\left[ 1+\eta _{v}-2a(\rho_i ) \right] \,,  \label{3.7a} \\
u_{10} &=&\vartheta_0 - \frac{\Delta_v}{2} - \frac{l}{\sqrt{2}\tau }\left[
2\eta _{v}a(\rho_i )- (\eta _{v}-1)\right] \,.  \label{3.7b}
\end{eqnarray}
where we introduced the quantities:
\begin{eqnarray*}
\Delta _{v} &=&\vartheta (0,\rho_i )-\vartheta (1,\rho_i )\,, \\
\eta _{v} &=&\left[ 1+\left( \frac{\tau \Delta _{v}}{2\sqrt{2}l}\right)
^{2}\right] ^{1/2}+\frac{\tau \Delta _{v}}{2\sqrt{2}l}\,.
\end{eqnarray*}
and $\rho_i$ is the corresponding value of the car density inside the
layers $\Im_{01}$ and $\Im_{10}$.

Expressions (\ref{3.7a}), (\ref{3.7b}) describe the regular dynamics of the
car cluster formation because the transition, for example, from the
free flow  to the synchronized phase  at a certain point $x$ is induced by
this transition at the nearest points. The dependence of the velocities
$u_{01}$ and $u_{10}$ on the local car density $\rho _{i}$
is illustrated in Fig.~\ref{F9}. The characteristic velocities
attained in this type motion can be estimated as
\[
\vartheta _{0}-u\sim \max \left\{ \vartheta _{0}\Delta /\rho _{c},\,l/\tau
\right\} \,,
\]
so, under the adopted assumptions the regular dynamics does not allow for
the sufficiently fast motion of the layers $\Im _{h}$ upstream.

\subsection{Noise-induced dynamics}

As the car density $\rho $ tends to the critical values $\rho _{c}$ or $
\rho _{c}+\Delta $ the value of $a(\rho )$ comes close to the boundaries $
a(\rho _{c})=1$ and $a(\rho _{c}+\Delta )=0$, and the point $h=1$ or $h=0$
becomes unstable, respectively. In this case the effect of the random
fluctuations $\xi (x,t)$ plays a substantial role. Namely, the phase
transition, for example, from the free flow to the synchronized motion
(for $\rho \approx \rho _{c}+\Delta $) is caused by the noise $\xi (x,t)$ and
equiprobably takes place at every point of the region wherein $\rho \approx
\rho _{c}+\Delta$ rather than is localized near the current position of
the layer $\Im _{01}$. Under these conditions the motion of the layers
$\Im _{h}$ can be qualitatively characterized by an extremely high velocity
in both the directions, which is illustrated in Fig.~\ref{F9} by dashed lines.

We note that the noise-induced motion, in contrast to the regular dynamics,
is to exhibit significant fluctuations in the displacement of the layer $\Im
_{h}$ as well as in its forms. This question is, however, a subject for
an individual study.

\subsection{Diffusion limited motion of vehicle clusters}

Let us, now, analyze the motion of a sufficiently large cluster that can
form on a highway when the initial car density or, what is the same, the
average car density $\bar{\rho}$ belongs to the metastable region, $\bar{
\rho}\in (\rho _{c},\,\rho _{c}+\Delta )$. The term ``sufficiently large''
means that the cluster dimension $L$ is assumed to be much greater than the
front distance $L^{*}$ looked through by drivers, so, they cannot look round
the congestion as a whole. Exactly, in this case a quasi-local description of
traffic flow similar to the differential equations~(\ref{3.1}), (\ref{3.2})
is justified.

\begin{figure}[b]
\centerline{\psfig{file=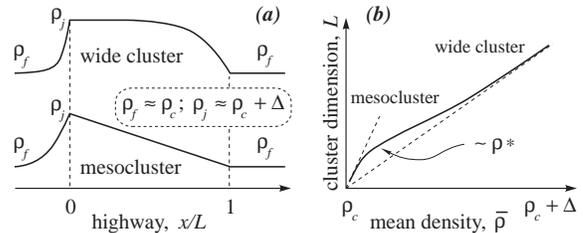}} \caption{The possible forms of the car
clusters and their dimension vs. the mean car density.} \label{F10}
\end{figure}

Converting to the frame $y=x-ut$ moving at the cluster velocity $u$, solving
Eq.~(\ref{3.1}) individually for the free flow and the synchronized phase, and
treating the layers $\Im _{h}$ as infinitely thin interfaces we get the
following conclusion. Within the frameworks of the given model the car
cluster moves upstream sufficiently fast, so, the motion of the layers $\Im
_{01}$ and $\Im _{10}$ is governed by the noise $\xi (x,t)$. In this case
the values of the car density at the layers $\Im _{01}$ and $\Im _{10}$
have to be $\rho _{j}\approx \rho _{c}+\Delta $ and $\rho _{f}\approx \rho
_{c}$, respectively. Thereby, the  cluster velocity $u$  is mainly
determined by the car redistribution governed by the diffusion type processes.
The latter feature is the reason why we refer to the cluster dynamics under
such conditions as to the diffusion limited motion. The transition region
${\cal L}_{01}$ between
practically the uniform free flow state and the cluster contains the
exponential increase of the vehicle density inside the free flow phase from
the value $\rho _{f}$ far from the ``interface'' $\Im _{01}$ up to $\rho
_{j}\approx \rho _{c}+\Delta $ at $\Im _{01}$,
\[
\rho =\rho _{f}+(\rho _{j}-\rho _{f})\exp \{q_{f}y\}\,,
\]
where $q_{f}=(\vartheta _{0}+\left| u\right| )/D\sim 1/L^{*}$ and the frame $
\{y\}$ is attached to the ``interface'' $\Im _{01}$. The transition region $
{\cal L}_{10}$ from the synchronized phase to the uniform free flow is to be
localized inside the car cluster. So, it is characterized by the
decrease in the vehicle density $\delta \rho \propto \exp \{q_{j}y\}$, where
$q_{j}=(\left| u\right| -\nu )/D$, and the vehicle free flow leaving the
cluster is uniform at all its points (Fig.~\ref{F10}$a$).

The cluster velocity is directly determined by the motion of the interface
$\Im _{01}$. Therefore, assuming also the cluster dimension $L$ large in
comparison with $L^{*}$, from Eq.~(\ref{3.1}) we get the expression of the
same form as the Lighthill--Whitham formula~(\ref{intro:4}) relating the
cluster velocity $u$ and the vehicle flux characteristics on the both sides
of the layer ${\cal L}_{01}$. Whence it follows that at the first
approximation:
\begin{equation}
u\approx -\nu \,,  \label{3.8}
\end{equation}
the value $q_{j}=0$, and the vehicle cluster is of the form shown in
Fig.~\ref{F10}$a$ under the name ``mesocluster''.
Assuming the total number of cars on the highway of length $L_{\text{rd}}$
fixed we get the expression for the mesocluster dimension $L$:
\begin{equation}
L=2L_{\text{rd}}\frac{\bar{\rho}-\rho _{c}}{\Delta }\,.
\label{3.9a}
\end{equation}

However, this result is justified only for sufficiently small values of $(
\bar{\rho}-\rho _{c})/\Delta \ll 1\,$, when the cluster dimension is not
too large, $Lq_{j}\ll 1$ (nevertheless, $L\gg L^{*}$). Exactly for this
reason we refer to such clusters as mesoscopic ones. In order to study the
opposite limit, $Lq_{j}\gg 1$, we have to take into account that the value
$\rho _{f}$ is not rigorously equal to $\rho _{c}$ but practically is the
root $\rho _{f}^{*}>\rho _{c}$ of the equation $u_{10}(\rho _{f}^{*})=-\nu $.
In this case the Lighthill--Whitham formula~(\ref{intro:4}) gives the
expression:
\[
u\simeq -\left[ \nu +(\vartheta _{0}+\nu )\frac{\rho _{f}^{*}-\rho _{c}}
{\Delta }\right]
\]
leading to the following estimates of the thickness $1/q_{j}$ of the
transition region ${\cal L}_{10}$:
\[
1/q_{j}\sim \frac{D\Delta }{(\vartheta _{0}+\nu )(\rho _{f}^{*}-\rho _{c})}
\sim L^{*}\frac{\Delta }{(\rho _{f}^{*}-\rho _{c})}\,.
\]
The form of such a wide cluster is shown in Fig.~\ref{F10}$a$, its dimension is:
\begin{equation}
L=L_{\text{rd}}\frac{\bar{\rho}-\rho _{c}}{\Delta }\,.  \label{3.9b}
\end{equation}
and the region of the mean car density corresponding to this limit is
specified by the inequality:
\begin{equation}
\frac{\rho ^{*}-\rho _{c}}{\Delta }\gg \frac{L^{*}}{L_{\text{rd}}} \frac{%
\Delta }{(\rho _{f}^{*}-\rho _{c})}\,.  \label{3.10}
\end{equation}

The resulting dependence of the cluster dimension on the mean car density
$\bar{\rho}$ is illustrated in Fig.~\ref{F10}{$b$}.

\section{Phase transition ``synchronized mode $\leftrightarrow$ jam''. Brief
discussion}
\label{sec:4}

In Sec.~\ref{sec:2} we have considered the phase transition between the
free flow and the synchronized mode. However, according to the
experimental data \cite{K98} there is an additional phase transition in
traffic flow regarded as the transition between the synchronized motion
and the jammed ``stop-and-go'' traffic. This transition occurs at
extremely high vehicle densities $\rho$ coming close to the limit value
$\rho _{0}$.

The present section briefly demonstrates that the developed model for the
driver behavior also predicts a similar phase transition at high car
densities. To avoid possible misunderstandings we, beforehand, point out
that the model in its present form cannot describe details of the
transition ``synchronized mode $\leftrightarrow$ jam'' because we have not
taken into account the delay in the driver response to variations in
headway. The latter is responsible for the formation of the
``stop-and-go'' pattern, so, to describe the jammed traffic on multilane
highways we, at least, should combine a governing equation for the order
parameter $h$ and a continuity equation similar to (\ref{3.2}),
(\ref{3.1}) with an equation for the car velocity relaxation similar to
(\ref{intro:2}). This question, however, is worthy of individual study.
Besides, the approximations used in Sec.~\ref{sec:3} to characterize the
synchronized mode at the car densities near $\rho_c$ do not hold here.

\begin{figure}
\centerline{\psfig{file=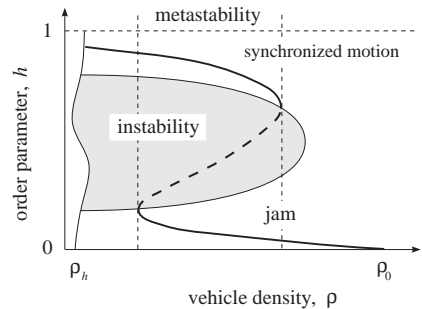,width=55 mm}} \caption{The instability
region and the $h(\rho)$-dependence describing the transition from the
synchronized (congested) phase to the heavy congested phase (a jam) in the
region of high car density.} \label{F11}
\end{figure}

In Sec.~\ref{sec:2} we have studied the dependence of the order parameter
$h$ on the car density ignoring the first term on the right-hand side of
Exp.~(\ref{2.7}) caused by the dangerous of lane changing. This assumption
is justified when the car density is not to high. In extremely dense
traffic flow, when the car density exceeds a certain value, $\rho >
\rho_h\lesssim \rho_0$, changing lanes becomes sufficiently dangerous and
the function $\Phi (h,v,\rho )$ describing the driver behavior is to
depend strongly on the vehicle density in this region. In addition, the
vehicle motion becomes sufficiently slow. Under such conditions the former
term on the right-hand side of expression~(\ref{2.7}) should be dominant
and, thus, $\partial\phi/\partial\rho >0$. Therefore, the stable vehicle
motion corresponding to $\partial \phi /\partial h>0$ matches the
decreasing dependence of the order parameter $h(\rho )$ on the vehicle
density $\rho $ for $\rho >\rho_h$. So, as the vehicle density $\rho$
increases the curve $h(\rho )$ can again go into the instability region
(in the $h\rho $-plane), which has to give rise to a jump from the
synchronized mode with greater values of the order parameter to a new
traffic state with its less values (Fig.~\ref{F11}). Obviously, this
transition between the two congested phases also exhibit the same
hysteresis as one describe in Sec.~\ref{sec:2}.

We identify the latter traffic state with the jammed vehicle motion.
Indeed, in extremely dense traffic lane change is practically depressed,
making the car ensembles at different lanes independent of one another.
So, in this case vehicle flow has to exhibit weak multilane correlations
and we should ascribe to it small values of the order parameter $h$. It
should be noted that the experimental single-vehicle data \cite{Last}
demonstrates strong correlations of variations in the traffic flux and the
car density for both the free flow and the ``stop-and-go'' motion. By
contrast, the synchronized mode is characterized by small values of the
cross-covariance between flow, speed, and density. In other words, for the
free flow and the ``stop-and-go'' motion the traffic flux
$j=\vartheta\rho$
should depends directly on the car density $\rho$, as it must in the
present model if we set $h=0$.

Finalizing the present section we point out that the given model treats
the jammed phase as a ``faster'' vehicle motion then the synchronized mode
at the {\it same} values of the order parameter. There is no contradiction
with the usual view on the synchronized mode as a high flux traffic state.
The latter corresponds to the traffic flow at the vehicle densities near
the phase transition ``free flow $\leftrightarrow$ synchronized mode''
rather than close to the limit value $\rho_0$. Besides, ordinary driver's
experience prompts that a highly dense traffic flow can be blocked at all
if one of the cars begin to change the lanes. Nevertheless, in order to
describe, at least qualitatively, the real features of the phase
transition ``synchronized mode $\leftrightarrow$ stop-and-go waves'' a
more sophisticated model is required. The present description only relates
it to the instability of the order parameter at high values of the vehicle
density.

Besides, the present analysis demonstrates also the nonmonotonic behavior
of the order parameter as the car density
increases even if we ignore the
hysteresis regions and focus our attention the stabel vehicle flow regions
only. It should be noted that a similar nonmonotonic dependence of the
lane change frequency on the car density as well as the platoon formation
has been found in the cellular automaton model for two-lane traffic
\cite{CA}.

\section{Closing remarks\label{sec:cr}}

Concluding the paper we recall the key points of the developed model.

We have proposed an original  macroscopic approach to the description of
multilane traffic flow based on an extended collection of the traffic flow
state variables. Namely, in addition to such characteristics as the car
density $\rho$ and mean velocity $v$ being actually the zeroth and first
moments of the ``one particle'' distribution function we introduce a new
variable $h$ called the ``order parameter''. It stands for the {\it
internal} correlations in the car motion along different lanes that are
due to the lane changing manoeuvres. The order parameter, in fact, allows
for the essentially ``many-body'' effects in the car interaction so it is
treated as an independent state variable.

Taking into account the general properties of the driver behavior we have
stated a governing equation for the order parameter.
Based on current
experimental data \cite{KR2,K98,KR1,KR3,Last} we have assume the
correlations in the car motion on multilane highways to be due to a small
group of ``fast'' drivers, i.e. the drivers who move substantially faster
than the statistically mean vehicle continuously overtaking other cars.
These ``fast'' cars, on one hand, increase individually the total rate of
vehicle flow but, on the other hand, make the accident danger greater and,
thus, cause the statistically mean driver to decrease the velocity. The
competition of the two effects depends on the car density and the mean
velocity and, as shown, can give rise to the traffic flow instability. It
turns out that the resulting dependence of the order parameter on the car
density describes in the same way the experimentally observed sequence of
phase transitions ``free flow $\leftrightarrow$ synchronized motion
$\leftrightarrow$ jam'' typical for traffic flow on highways \cite{K98}.
Besides, we have shown that both these transitions should be of the first
order type and exhibit hysteresis, matching the experimental data
\cite{K98,KR1,KR3}. The synchronized mode is characterized by a large
value of the order parameter, whereas the free flow and the jam match its
small values. The latter feature enables us to treat the jam as a phase
comprising the vehicle flows at different lane with weak mutual
interaction because of the lane changing being depressed.

In order to illustrate the characteristic features of the car clusters that
self-organizing under these conditions we have considered a simple model
dealing only with the evolution of the car density and the order
parameter. In particular, it is shown that in the steady state
the car density inside the cluster and the free flow being in equilibrium
with the cluster, as well as the velocity at which the cluster moves
upstream are fixed and determined by the basic properties of the traffic
flow. On the contrary, the size of the car cluster depends on the initial
conditions.

Finally, we would like to underline that the developed model takes into
account only one effect causing the traffic flow instability. The other,
the delay in the driver control over the headway, seems to be responsible
for the ``stop-and-go'' waves in the jammed phase (for a review of the
continuum description of this phenomena see, e.g., \cite{T99a,T99b}). So,
combining the two approaches into one model it enables detail description
of a wide class of phenomena occurring in the transitions from free flow
to the heavy congested phase on highways. In this way the order parameter
model could describe also the formation of a local jam on a highway whose
boundaries comprise both of the phase transitions. In the present form it
fails to do this because the free flow and the jammed traffic are
characterized by small values of the order parameter.

Concerning a possible derivation of the order parameter model from the
gas-kinetic theory we note that the appearance of the ``fast'' driver
platoons demonstrates a substantial deviation of the car distribution
function from the monotonic quasi-equilibrium form. So, to construct an
adequate  system of equations dealing with the moments of the distribution
function a more sophisticated approximation is required.

\section*{Acknowledgments}

One of us, I.\,A.~Lubashevsky, would like to acknowledge the hospitality of
Physics Department of Rostock University during the stay at which this work
was carried out.


\begin{references}
\bibitem{LM00} To obtain estimate (\protect\ref{3.5}) we need a special analysis and
its presentation would substantially overload the given paper. Moreover, the amplitude
of the random fluctuations  does not enter the obtained formulae, so, we refer
this question to our forthcoming paper.
\end{references}
\end{document}